\documentclass[10pt]{article}
\usepackage{graphicx}
\usepackage{amsmath}
\usepackage{amssymb}
\usepackage{caption2}
\begin{document}
\bibliographystyle{prsty}
\begin{center}
{\large {\bf \sc{Analysis of  the bottom-charm baryon states  with QCD sum rules }}} \\[2mm]
Zhi-Gang Wang \footnote{E-mail: zgwang@aliyun.com.  }, Qi Xin     \\
 Department of Physics, North China Electric Power University, Baoding 071003, P. R. China
\end{center}

\begin{abstract}
In this work, we  explore the ground state mass spectrum of the bottom-charm baryon states in the framework of
 the QCD sum rules in a comprehensive way. In calculations,   we distinguish  the contributions of the positive-parity and negative-parity baryon  states  unambiguously,
and  investigate the energy scale dependence of the QCD sum rules in details and observe that the acceptable energy scales are $\mu=2.0\sim 2.8\,\rm{GeV}$, while the best energy scale is $\mu=2.0\,\rm{GeV}$.  The predictions can be confronted to the experimental data in the future and shed light on the QCD low energy dynamics.
\end{abstract}



\section{Introduction}

In 2017, the LHCb collaboration observed the doubly charmed baryon state  $\Xi_{cc}^{++}$ in the $\Lambda_c^+ K^- \pi^+\pi^+$ mass spectrum  \cite{LHCb-Xicc}.
The observation of the $\Xi_{cc}^{++}$ provides us with the crucial experimental input on the strong correlation between the two charm valence quarks, which maybe shed light on the spectrum of the doubly-heavy  baryon states and multiquark states,  and has attracted  great attentions in the hadron physics.
 The $\Xi^{++}_{cc}$ contains two charm valence quarks, which requires the $cc$-diquark should  only be flavor symmetric, we can construct the diquark operators  $\varepsilon^{ijk}c^T_j C\gamma_\mu c_k$ and $\varepsilon^{ijk}c^T_j C\sigma_{\mu\nu} c_k$ to satisfy the Fermi-Dirac statistics, where the $i$, $j$ and $k$ are color indexes; while in the  case that there are  two different heavy-flavor valence quarks, the $bc$-diquark could be either flavor symmetric or antisymmetric, we can construct the diquark operators $\varepsilon^{ijk}b^T_j C\gamma_\mu c_k$,  $\varepsilon^{ijk}b^T_j C\sigma_{\mu\nu} c_k$, $\varepsilon^{ijk}b^T_j C\gamma_5 c_k$, they all satisfy the Fermi-Dirac statistics.  In this work, we would focus on the latter ones, the baryon states having flavor constituents $bcq$ with the light quarks $q=u$, $d$, or $s$. As the $c$-quark is not heavy enough, we can also construct the $\varepsilon^{ijk}q^T_j C\gamma_\mu c_k$,  $\varepsilon^{ijk}q^T_j C\sigma_{\mu\nu} c_k$, $\varepsilon^{ijk}q^T_j C\gamma_5 c_k$ diquark operators to study the bottom-charm baryon states.

The bottom-charm baryon states have been studied in the framework of the
(simple) potential models \cite{Bagan-PM+QCDSR,PM-Shah,PM-Weng} (\cite{Karliner-Simple-QM}),
 Feynman-Hellmann plus semi-empirical formulas \cite{Lichtenberg-Feynman},
relativistic quasipotential quark model \cite{Ebert-bc,Ebert-bc-2002},
potential model plus  non-relativistic  QCD sum rules \cite{NRQCDSR-Kiselev},
MIT Bag model \cite{MITBag-LiXQ},
  variational ansatz based on the non-relativistic potential model \cite{Var-Albertus,Var-Roberts},
Bethe-Salpeter equation \cite{BSE-GuoXH,BSE-WangGL},
 QCD sum rules \cite{ZhangJR-Doubly}, Lattice QCD \cite{LattQCD-doubly}, etc. The predicted masses for the  ground state bottom-charm baryon states $\Xi_{bc}$ and $\Omega_{bc}$ are about $6.8\sim7.0\,\rm{GeV}$ and $6.9\sim7.1\,\rm{GeV}$, respectively.

The LHCb collaboration searched for the doubly-heavy baryon $\Xi_{bc}^0$ in the
$D^0pK^-$ mass spectrum and observed no significant signal in the invariant
mass range $6.7\sim 7.2 \,\rm{GeV}$ \cite{LHCb-Xcc-No-2009}, and searched for the doubly-heavy baryons $\Omega_{bc}^0$ and $\Xi_{bc}^0$
in the $\Lambda_c^+\pi^-$ and $\Xi_c^+\pi^-$ mass spectrum and observed no
significant excess in the invariant mass range $6.7\sim 7.3 \,\rm{GeV}$ \cite{LHCb-Xcc-No-2104}. More theoretical and experimental works are still needed to
explore the bottom-charm baryon states.

In Refs.\cite{WangHHbaryon-4-P,WangH-HHbaryon-6-4-P,WangH-HHbaryon-6-5-N,WangHbaryon-6-PN,WZG-EPJC-cc-baryon,WZG-AAPPS},  we distinguish the contributions of the positive parity and negative parity baryon states explicitly, and investigate  the  heavy, doubly-heavy and triply-heavy baryon states  with the QCD sum rules systematically.
The works \cite{WangHHbaryon-4-P,WangH-HHbaryon-6-4-P,WangH-HHbaryon-6-5-N,WangHbaryon-6-PN} were finished before the observation of the $\Xi^{++}_{cc}$ by the LHCb collaboration \cite{LHCb-Xicc}, while the works \cite{WZG-EPJC-cc-baryon,WZG-AAPPS} were finished after the discovery of the $\Xi_{cc}^{++}$, where we have taken account of the new experimental  input to explore the doubly-heavy and triply-heavy baryon states. Now we extend our previous works to investigate the bottom-charm baryon states in the framework of the QCD sum rules in details by carrying out the operator product expansion  up to  the vacuum condensates of dimension $7$, just like in Ref.\cite{WZG-EPJC-cc-baryon}.

 The article is arranged as:
  we obtain  the QCD sum rules for the masses and pole residues of the bottom-charm baryon states  in Sect.2;  in Sect.3, we present the numerical results and discussions; and Sect.4 is reserved for our
conclusion.

\section{QCD sum rules for  the bottom-charm baryon states}

Now we write down  the  correlation functions $\Pi(p)$ and $\Pi_{\mu\nu}(p)$  in the QCD sum rules,
\begin{eqnarray}
\Pi(p)&=&i\int d^4x e^{ip \cdot x} \langle0|T\left\{J(x)\bar{J}(0)\right\}|0\rangle \, , \nonumber\\
\Pi_{\mu\nu}(p)&=&i\int d^4x e^{ip \cdot x} \langle0|T\left\{J_{\mu}(x)\bar{J}_{\nu}(0)\right\}|0\rangle \, ,
\end{eqnarray}
where the currents $\bar{J}(x)=J^\dagger(x) \gamma^0$, $\bar{J}_\mu(x)=J_\mu^\dagger(x) \gamma^0$, $J(x)=J^{S}(x)$, $J^{A}(x)$, $\eta^{S}(x)$, $\eta^{A}(x)$, $J_\mu(x)= J_\mu^{A}(x)$, $\eta_\mu^{A}(x)$,
\begin{eqnarray}
 J^{S}(x)&=&   \varepsilon^{ijk}\, b^T_i(x) C\gamma_5 c_j(x) \, q_{k}(x) \, ,\nonumber \\
 J^{A}(x)&=&   \varepsilon^{ijk} \, b^T_i(x) C\gamma_\mu c_j(x)\gamma_5\gamma^\mu q_{k}(x) \, ,\nonumber \\
 J_\mu^{A}(x)&=&   \varepsilon^{ijk}\,  b^T_i(x) C\gamma_\mu c_j(x)\,  q_{k}(x) \, ,\nonumber \\
\eta^{S}(x)&=&   \varepsilon^{ijk} \, q^T_i(x) C\gamma_5 c_j(x) \, b_{k}(x) \, ,\nonumber \\
 \eta^{A}(x)&=&   \varepsilon^{ijk} \, q^T_i(x) C\gamma_\mu c_j(x)\gamma_5\gamma^\mu b_{k}(x) \, ,\nonumber \\
 \eta_\mu^{A}(x)&=&   \varepsilon^{ijk}\,  q^T_i(x) C\gamma_\mu c_j(x)\,  b_{k}(x) \, ,
\end{eqnarray}
$q=u$, $d$, $s$, the $i$, $j$, $k$ are color indexes,   the superscripts $S$ and $A$ denote the scalar and axialvector diquark operators  $\varepsilon^{ijk}  b^T_i C\gamma_5 c_j$ ($\varepsilon^{ijk}  q^T_i C\gamma_5 c_j$) and $\varepsilon^{ijk}  b^T_i C\gamma_\mu c_j$ ($\varepsilon^{ijk}  q^T_i C\gamma_\mu c_j$), respectively.

The   currents $J(0)$ and $J_{\mu}(0)$ couple potentially to the spin-party $J^P={\frac{1}{2}}^\pm$ and ${\frac{1}{2}}^\mp$,  ${\frac{3}{2}}^\pm$ bottom-charm baryon    states $B_{\frac{1}{2}}^{\pm}$ and $B_{\frac{1}{2}}^{\mp}$, $B_{\frac{3}{2}}^{\pm}$, respectively,
\begin{eqnarray}
\langle 0| J (0)|B_{\frac{1}{2}}^{+}(p)\rangle &=&\lambda^{+}_{\frac{1}{2}}\,  U^{+}(p,s) \, , \nonumber \\
\langle 0| J (0)|B_{\frac{1}{2}}^{-}(p)\rangle &=&\lambda^{-}_{\frac{1}{2}}\,i\gamma_5  U^{-}(p,s) \, , \nonumber \\
\langle 0| J_{\mu} (0)|B_{\frac{1}{2}}^{-}(p)\rangle &=&f^{-}_{\frac{1}{2}}\,p_\mu \,U^{-}(p,s) \, , \nonumber \\
\langle 0| J_{\mu} (0)|B_{\frac{1}{2}}^{+}(p)\rangle &=&f^{+}_{\frac{1}{2}}\,p_\mu i\gamma_5\,U^{+}(p,s) \, , \nonumber \\
\langle 0| J_{\mu} (0)|B_{\frac{3}{2}}^{+}(p)\rangle &=&\lambda^{+}_{\frac{3}{2}}\, U^{+}_\mu(p,s) \, ,\nonumber \\
\langle 0| J_{\mu} (0)|B_{\frac{3}{2}}^{-}(p)\rangle &=&\lambda^{-}_{\frac{3}{2}}\,i\gamma_5 U^{-}_\mu(p,s) \, ,
\end{eqnarray}
 where the  $\lambda^{\pm}_{\frac{1}{2}}$, $\lambda^{\pm}_{\frac{3}{2}}$ and $f^{\pm}_{\frac{1}{2}}$  are the pole residues, the
 $U^\pm(p,s)$ and $U^{\pm}_\mu(p,s)$ are the Dirac and Rarita-Schwinger spinors, respectively \cite{WangHHbaryon-4-P,WangH-HHbaryon-6-4-P,WangH-HHbaryon-6-5-N,WangHbaryon-6-PN,WZG-EPJC-cc-baryon,WZG-AAPPS}.

 At the hadron side of the correlation functions $\Pi(p)$ and $\Pi_{\mu\nu}(p)$, we  isolate the pole terms of the lowest bottom-charm baryon states with the positive parity and negative parity, and acquire the hadron representation,
 \begin{eqnarray}
 \Pi(p) & = & {\lambda^{+}_{\frac{1}{2}}}^2  {\!\not\!{p}+ M_{+} \over M_{+}^{2}-p^{2}  } +  {\lambda^{-}_{\frac{1}{2}}}^2  {\!\not\!{p}- M_{-} \over M_{-}^{2}-p^{2}  } +\cdots  \, ,  \nonumber\\
 &=&\Pi^1_{\frac{1}{2}}(p^2)\!\not\!{p} +\Pi^0_{\frac{1}{2}}(p^2)\, ,
 \end{eqnarray}
 \begin{eqnarray}
 \Pi_{\mu\nu}(p) & = & {\lambda^{+}_{\frac{3}{2}}}^2  {\!\not\!{p}+ M_{+} \over M_{+}^{2}-p^{2}  } \left(- g_{\mu\nu}+\cdots
\right)
+  {\lambda^{-}_{\frac{3}{2}}}^2  {\!\not\!{p}- M_{-} \over M_{-}^{2}-p^{2}  } \left(- g_{\mu\nu}+\cdots \right)  +\cdots \nonumber \\
 &=&\left[\Pi^1_{\frac{3}{2}}(p^2)\!\not\!{p} +\Pi^0_{\frac{3}{2}}(p^2)\right] \left(-g_{\mu\nu} \right)+\cdots\, .
\end{eqnarray}
In this work, we choose the tensor/spin structures $1$, $\!\not\!{p}$, $g_{\mu\nu}$ and $\!\not\!{p}g_{\mu\nu}$ to investigate the bottom-charm baryon states with the  $J^P={\frac{1}{2}}^+$ and ${\frac{3}{2}}^+$, respectively.
We obtain the hadronic  spectral densities  through the dispersion relation,
\begin{eqnarray}
\frac{{\rm Im}\Pi^1_{j}(s)}{\pi}&=&{\lambda^{+}_{j}}^2 \delta\left(s-M_{+}^2\right)+{\lambda^{-}_{j}}^2 \delta\left(s-M_{-}^2\right)\, , \nonumber\\
                                &=&\rho^1_{j,H}(s)\, ,\nonumber\\
\frac{{\rm Im}\Pi^0_{j}(s)}{\pi}&=&M_{+}{\lambda^{+}_{j}}^2 \delta\left(s-M_{+}^2\right)-M_{-}{\lambda^{-}_{j}}^2 \delta\left(s-M_{-}^2\right)\, , \nonumber\\
&=& \rho^0_{j,H}(s) \, ,
\end{eqnarray}
where $j=\frac{1}{2}$ and $\frac{3}{2}$, we introduce  the subscript $H$ to represent the hadron side. Then we introduce the weight function $\exp\left(-\frac{s}{T^2}\right)$ to obtain the QCD sum rules at  the hadron side,
\begin{eqnarray} \label{QCDSR-H}
\int_{(m_b+m_c)^2}^{s_0}ds \left[\sqrt{s}\rho^1_{j,H}(s)+\rho^0_{j,H}(s)\right]\exp\left( -\frac{s}{T^2}\right)
&=&2M_{+}{\lambda^{+}_{j}}^2\exp\left( -\frac{M_{+}^2}{T^2}\right) \, ,
\end{eqnarray}
where the $s_0$ are the continuum threshold parameters and the $T^2$ are the Borel parameters.
We choose the  special combinations $\sqrt{s}\rho^1_{j,H}(s)+\rho^0_{j,H}(s)$ in Eq.\eqref{QCDSR-H} to pick up  the  contributions  of the positive  parity baryon  states from that of the negative  parity baryon  states unambiguously.

 At the QCD side of the correlation functions $\Pi(p)$ and $\Pi_{\mu\nu}(p)$, we carry out  the operator product expansion  up to the vacuum condensates of dimension $7$.  There are two heavy quark lines and one  light quark line, if each heavy quark line emits a gluon and each light quark line gives  a quark pair, we obtain an  operator $g_s^2GG\bar{q}q$, which is of dimension $7$,   we should take  account of the vacuum condensates  up to dimension $7$ at least.   We take the truncations $n\leq 7$ and $k\leq 1$  consistently according to our previous works,
the operators of the orders $\mathcal{O}( \alpha_s^{k})$ with $k> 1$ are  discarded \cite{WZG-EPJC-cc-baryon,WangHuangTao-3900,Wang-tetra-formula,WangZG-molecule}.

Now let us  take the quark-hadron duality below the continuum thresholds  $s_0$ and again introduce the weight function $\exp\left(-\frac{s}{T^2}\right)$ to obtain  the QCD sum rules:
\begin{eqnarray}\label{QCDSR}
2M_{+}{\lambda^{+}_{j}}^2\exp\left( -\frac{M_{+}^2}{T^2}\right)
&=& \int_{(m_b+m_c)^2}^{s_0}ds \left[\sqrt{s}\rho^1_{j,QCD}(s)+\rho^0_{j,QCD}(s)\right]\exp\left( -\frac{s}{T^2}\right)\, ,
\end{eqnarray}
where the $\rho^1_{j,QCD}(s)$ and $\rho^0_{j,QCD}(s)$ are the QCD spectral densities corresponding to the hadronic  spectral densities $\rho^1_{j,H}(s)$ and $\rho^0_{j,H}(s)$, respectively. Their explicit expressions are neglected for simplicity, the interested readers can acquire them via contracting the corresponding author via E-mail.

We differentiate  Eq.\eqref{QCDSR} in regard  to  $\tau=\frac{1}{T^2}$, then eliminate the
 pole residues $\lambda^{+}_{j}$ and acquire  the QCD sum rules for
 the masses of the bottom-charm baryon states,
 \begin{eqnarray}
 M^2_{+} &=& \frac{-\frac{d}{d \tau}\int_{(m_b+m_c)^2}^{s_0}ds \,\left[\sqrt{s}\,\rho^1_{j,QCD}(s)+\,\rho^0_{j,QCD}(s)\right]\exp\left(- \tau s\right)}{\int_{(m_b+m_c)^2}^{s_0}ds \left[\sqrt{s}\,\rho_{j,QCD}^1(s)+\,\rho_{j,QCD}(s)\right]\exp\left( -\tau s\right)}\, .
\end{eqnarray}

\section{Numerical results and discussions}
We adopt  the conventional  values of the vacuum condensates $\langle
\bar{q}q \rangle=-(0.24\pm 0.01\, \rm{GeV})^3$,   $\langle
\bar{q}g_s\sigma G q \rangle=m_0^2\langle \bar{q}q \rangle$,
$m_0^2=(0.8 \pm 0.1)\,\rm{GeV}^2$, $\langle\bar{s}s \rangle=(0.8\pm0.1)\langle\bar{q}q \rangle$, $\langle\bar{s}g_s\sigma G s \rangle=m_0^2\langle \bar{s}s \rangle$,  $\langle \frac{\alpha_s
GG}{\pi}\rangle=0.012\pm 0.004\,\rm{GeV})^4 $    at the energy scale  $\mu=1\, \rm{GeV}$
\cite{SVZ79,PRT85,ColangeloReview}, and choose the $\overline{MS}$ masses $m_{c}(m_c)=(1.275\pm0.025)\,\rm{GeV}$, $m_{b}(m_b)=(4.18\pm0.03)\,\rm{GeV}$, $m_s(\mu=2\,\rm{GeV})=0.095 \pm 0.005\,\rm{GeV}$ from the Particle Data Group \cite{PDG}, and set $m_q=m_u=m_d=0$.
In addition, we take  account of the energy-scale dependence of  the input parameters,
\begin{eqnarray}
\langle\bar{q}q \rangle(\mu)&=&\langle\bar{q}q \rangle({\rm 1GeV})\left[\frac{\alpha_{s}({\rm 1GeV})}{\alpha_{s}(\mu)}\right]^{\frac{12}{33-2n_f}}\, , \nonumber\\
\langle\bar{s}s \rangle(\mu)&=&\langle\bar{s}s \rangle({\rm 1GeV})\left[\frac{\alpha_{s}({\rm 1GeV})}{\alpha_{s}(\mu)}\right]^{\frac{12}{33-2n_f}}\, , \nonumber\\
 \langle\bar{q}g_s \sigma Gq \rangle(\mu)&=&\langle\bar{q}g_s \sigma Gq \rangle({\rm 1GeV})\left[\frac{\alpha_{s}({\rm 1GeV})}{\alpha_{s}(\mu)}\right]^{\frac{2}{33-2n_f}}\, , \nonumber\\
 \langle\bar{s}g_s \sigma Gs \rangle(\mu)&=&\langle\bar{s}g_s \sigma Gs \rangle({\rm 1GeV})\left[\frac{\alpha_{s}({\rm 1GeV})}{\alpha_{s}(\mu)}\right]^{\frac{2}{33-2n_f}}\, , \nonumber\\
 m_Q(\mu)&=&m_Q(m_Q)\left[\frac{\alpha_{s}(\mu)}{\alpha_{s}(m_Q)}\right]^{\frac{12}{33-2n_f}} \, ,\nonumber\\
 m_s(\mu)&=&m_s({\rm 2GeV})\left[\frac{\alpha_{s}(\mu)}{\alpha_{s}({\rm 2GeV})}\right]^{\frac{12}{33-2n_f}} \, ,\nonumber\\
\alpha_s(\mu)&=&\frac{1}{b_0t}\left[1-\frac{b_1}{b_0^2}\frac{\log t}{t} +\frac{b_1^2(\log^2{t}-\log{t}-1)+b_0b_2}{b_0^4t^2}\right]\, ,
\end{eqnarray}
 where $q=u$, $d$,
   where $t=\log \frac{\mu^2}{\Lambda^2}$, $b_0=\frac{33-2n_f}{12\pi}$, $b_1=\frac{153-19n_f}{24\pi^2}$, $b_2=\frac{2857-\frac{5033}{9}n_f+\frac{325}{27}n_f^2}{128\pi^3}$,  $\Lambda=210\,\rm{MeV}$, $292\,\rm{MeV}$  and  $332\,\rm{MeV}$ for the flavors  $n_f=5$, $4$ and $3$, respectively  \cite{PDG,Narison-mix}, and evolve all the input parameters to the ideal energy scales   $\mu$ to extract the masses of the
   bottom-charm  baryon states with the flavor number $n_f=5$.

 The doubly-charm, doubly-bottom, hidden-charm, hidden-bottom tetraquark  states $X$, $Y$, $Z$ and pentaquark  states $P$
are characterized by the effective heavy quark masses ${\mathbb{M}}_Q$   and the virtuality
$V=\sqrt{M^2_{X/Y/Z/P}-(2{\mathbb{M}}_Q)^2}$, we choose the  energy scales  $\mu=V=\sqrt{M^2_{X/Y/Z}-(2{\mathbb{M}}_Q)^2}$, which
 works well \cite{Wang-tetra-formula,WangZG-molecule}. The updated values are ${\mathbb{M}}_c=1.82\,\rm{GeV}$ and ${\mathbb{M}}_b=5.17\,\rm{GeV}$ for the diquark-antidiquark type (diquark-diquark-antiquark type) hidden-charm and hidden-bottom tetraquark (pentaquark) states, respectively \cite{WZG-PRD-cc-spectrum,WZG-EPJC-bb-spectrum}.

For the lowest doubly-charmed baryon state $\Xi_{cc}^{++}$, the only doubly-heavy baryon state observed experimentally up to know, the updated  mass $M_{\Xi_{cc}^{++}}=3621.55 \pm 0.23 \pm 0.30  \,\rm{MeV}$ \cite{LHCb-Xicc-update}, which is smaller than $2 {\mathbb{M}}_c=3.64\,\rm{GeV}$, the energy scale formula $\mu=\sqrt{M^2_{\Xi^{++}_{cc}}-(2{\mathbb{M}}_c)^2}$ is failed to work. In the present work, if the energy scale formula works, the lowest mass,
\begin{eqnarray}
M_{\frac{1}{2}}&=& \sqrt{\mu^2+({\mathbb{M}}_b+{\mathbb{M}}_c)^2} \nonumber\\
               &\geq&\sqrt{(1\rm{GeV})^2+({\mathbb{M}}_b+{\mathbb{M}}_c)^2} \nonumber\\
               &=&7.06\,\rm{GeV}\, .
\end{eqnarray}
On the other hand, if the energy scale formula is also failed to work, just like in the case of the $\Xi_{cc}^{++}$, we expect the mass $M_{\frac{1}{2}}\leq {\mathbb{M}}_b+{\mathbb{M}}_c=6.99\,\rm{GeV}$.

\begin{figure}
 \centering
 \includegraphics[totalheight=7cm,width=10cm]{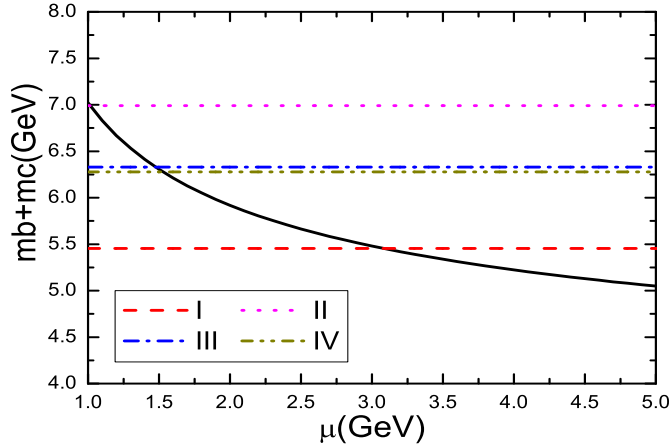}
 \caption{ The values of the $m_b(\mu)+m_c(\mu)$   with variations  of the energy scales $\mu$, where the I, II, III and IV denote the  $m_b(m_b)+m_c(m_c)$,
 ${\mathbb{M}}_b+{\mathbb{M}}_c$, $M_{B_c^*}$ and $\frac{M_{J/\psi}+M_{\Upsilon}}{2}$, respectively.  }\label{QuarkMass}
\end{figure}

In Fig.\ref{QuarkMass}, we plot the values of the $m_b(\mu)+m_c(\mu)$   with variations  of the energy scales $\mu$ of the QCD spectral densities. From the figure, we can see clearly that the values decrease monotonically and quickly with the increase of the energy scales, while the interval of the integral $(m_b+m_c)^2-s_0$ of the variable $ds$  increases monotonically and quickly with the increase of the energy scales $\mu$. Larger interval of the integral leads to more stable QCD sum rules, and we expect to choose larger energy scales, and it is better to choose the energy scales between $\mu=1.5\,\rm{GeV}$ and $3.0\,\rm{GeV}$.

At the energy scale about $\mu=1.5\,\rm{GeV}$, the value of the $m_b(\mu)+m_c(\mu)$ equals  to $M_{B_c^*}$ and $\frac{M_{J/\psi}+M_{\Upsilon}}{2}$ approximately,  the energy scale
$\mu=1.5\,\rm{GeV}$ is too small to obtain  reliable QCD sum rules. On the other hand, at the energy scale about $\mu=3.0\,\rm{GeV}$,  $m_b(\mu)+m_c(\mu)\approx m_b(m_b)+m_c(m_c)$, such an energy is too large to make reliable predictions, as in the QCD sum rules even for the fully-heavy bottom tetraquark states, the energy scale is only as large as $\mu=3.1\,\rm{GeV}$ \cite{WZG-QQQQ-EPJC,WZG-DZY-APPB}.

In Ref.\cite{WZG-EPJC-cc-baryon},  we choose the typical energy scales
$\mu=1\,\rm{GeV}$ for the doubly-charmed baryon states $\Xi_{cc}$,  $\Xi^*_{cc}$, $\Omega_{cc}$,  $\Omega^*_{cc}$ and  $\mu=2.2\,\rm{GeV}$
 for the doubly-bottom baryon states $\Xi_{bb}$,  $\Xi^*_{bb}$, $\Omega_{bb}$ and  $\Omega^*_{bb}$, and reproduce the experimental value of the mass of the $\Xi_{cc}^{++}$ satisfactorily. However, we should bear in mind that the energy scale $\mu=2.2\,\rm{GeV}$ is the lower bound, as emphasized in Ref.\cite{WZG-EPJC-cc-baryon}. Analogously, we expect to choose  $\mu\sim 2.2\,\rm{GeV}$ in the present work.

In Ref.\cite{WZG-AAPPS}, we re-explore the mass spectrum of the ground state triply-heavy baryon states with the QCD sum rules in details  and preform  a novel  analysis, and observe the optimal energy scales of the QCD spectral densities of the
$ccc({\frac{3}{2}}^+)$,   $ccb({\frac{3}{2}}^+)$,    $ccb({\frac{1}{2}}^+)$,  $bbc({\frac{3}{2}}^+)$,   $bbc({\frac{1}{2}}^+)$ and  $bbb({\frac{3}{2}}^+)$ are $\mu=1.2\,\rm{GeV}$, $2.1\,\rm{GeV}$,  $2.0\,\rm{GeV}$, $2.2\,\rm{GeV}$,  $2.2,\rm{GeV}$ and $2.5\,\rm{GeV}$, respectively. The optimal energy scales for the $QQQ^\prime$ baryon states are about $2.0\sim2.2\,\rm{GeV}$.

In Refs.\cite{WZG-QQQQ-EPJC,WZG-DZY-APPB}, we give detailed discussions on how to choose the pertinent energy scales of the QCD spectral densities for the diquark-antidiquark type fully-heavy tetraquark states, and obtain the optimal energy scales $\mu=2.0\,\rm{GeV}$ and $3.1\,\rm{GeV}$ for the   $cc\bar{c}\bar{c}$ and $bb\bar{b}\bar{b}$, respectively. In this scheme, we can account for the experimental data on the fully-charm tetraquark structures from the LHCb collaboration in a satisfactory way \cite{WZG-4Q-CPC}.

In Ref.\cite{WZG-NPB-5Q}, we investigate  the diquark-diquark-antiquark type fully-heavy pentaquark states with the spin-parity $J^P={\frac{1}{2}}^-$ in the framework of  the QCD sum rules in details, we choose the   typical energy scales $\mu=m_c(m_c)\, ({\rm i.e.}\, 1.275\,\rm{GeV})$ and $2.8\,\rm{GeV}$ to extract the  masses of the
 fully-heavy pentaquark states $cccc\bar{c}$ and $bbbb\bar{b}$, respectively.

In Ref.\cite{WZG-6Q}, we construct the diquark-diquark-diquark type vector six-quark currents to investigate the vector and scalar hexaquark states  in the framework of the QCD sum rules and choose the  typical energy scales $\mu=m_c(m_c)\, ({\rm i.e.}\, 1.275\,\rm{GeV})$ and $2.8\,\rm{GeV}$ to extract the  masses of the
 fully-heavy hexaquark states $cccccc$ and $bbbbbb$, respectively. In Refs.\cite{WZG-NPB-5Q,WZG-6Q}, we acquire every flat Borel platforms.

In summary,  we can obtain the conclusion tentatively that  the acceptable  energy scales of the bottom-charm baryon states are  $\mu=2.0-2.8\,\rm{GeV}$.
In the present work, we choose the typical energy scales $\mu=1.5\,\rm{GeV}$, $2.0\,\rm{GeV}$ and $2.5\,\rm{GeV}$ tentatively, and try to make suitable predictions.

In the following, we define the contributions of the different terms in the operator product expansion $D(n)$ and the pole contributions ($\rm{PC}$),
\begin{eqnarray}
D(n)&=& \frac{  \int_{(m_b+m_c)^2}^{s_0} ds\,\rho_{n}(s)\,\exp\left(-\frac{s}{T^2}\right)}{\int_{(m_b+m_c)^2}^{s_0} ds \,\rho(s)\,\exp\left(-\frac{s}{T^2}\right)}\, ,
\end{eqnarray}
\begin{eqnarray}
{\rm PC}&=& \frac{  \int_{(m_b+m_c)^2}^{s_0} ds\,\rho(s)\,
\exp\left(-\frac{s}{T^2}\right)}{\int_{(m_b+m_c)^2}^{\infty} ds \,\rho(s)\,\exp\left(-\frac{s}{T^2}\right)}\, ,
\end{eqnarray}
where the $\rho_{n}(s)$ are the QCD spectral densities involving  the vacuum condensates of dimension $n$, and the total spectral densities
$\rho(s)=\sqrt{s}\rho^1_{QCD}(s)+ \rho^0_{QCD}(s)$.

Now, let us estimate the mass of the lowest bottom-charm baryon state,
\begin{eqnarray}
M_{\frac{1}{2}}&=&M_{\Xi_{cc}}+M_{B_c^*}-M_{J/\psi}=6.85\,\rm{GeV} \, ,
\end{eqnarray}
which is consistent with the assumption $M_{\frac{1}{2}}\leq {\mathbb{M}}_b+{\mathbb{M}}_c=6.99\,\rm{GeV}$, and we have taken the masses $M_{B_c^*}=6.33\,\rm{GeV}$   from the CMS  Collaboration \cite{CMS-BcV} and
 $M_{J/\psi}=3.0969\,\rm{GeV}$ from the Particle Data Group \cite{PDG}, the energy scale formula is failed to work indeed. The value $6.85\,\rm{GeV}$ lies below the diquark-antidiquark type axialvector bottom-charm tetraquark states based on the  QCD sum rules, $M_{Z_{\bar{b}c}(1^{+-})}=7.30\pm0.08\,\rm{GeV}$ and $M_{Z_{\bar{b}c}(1^{++})}=7.31\pm0.08\,\rm{GeV}$, where the pertinent energy scales are  $\mu=2.10\,\rm{GeV}$ and $2.15\,\rm{GeV}$ respectively from the energy scale formula \cite{WZG-EPL-Bc}.

In calculations, we choose the continuum threshold parameters $\sqrt{s_0}=M_{gr}+ (0.5\sim0.7)\,\rm{GeV}$ as a constraint,  where the subscript $gr$ denotes the ground states or the lowest states. We search for the  Borel parameters $T^2$ and continuum threshold
parameters $s_0$  to satisfy   the  two basic criteria:  Pole dominance  and Convergence of the operator product expansion via trial and error.

The resulting Borel parameters, continuum threshold parameters  and  pole contributions of the ground states  are shown  explicitly in Tables \ref{Borel-mass-pole}-\ref{Borel-mass-pole-2}.
From the tables, we can see explicitly that the pole contributions are about $(50-75)\%$, the pole  dominance at the hadron side is satisfied very good.  At the QCD side, we observe that the main contributions come from the perturbative terms, the contributions of the highest dimensional vacuum condensates $\langle\bar{q}q\rangle\langle\frac{\alpha_sGG}{\pi}\rangle$ are much less than one percent, the operator product expansion converges also very good.

\begin{figure}
 \centering
 \includegraphics[totalheight=6cm,width=7cm]{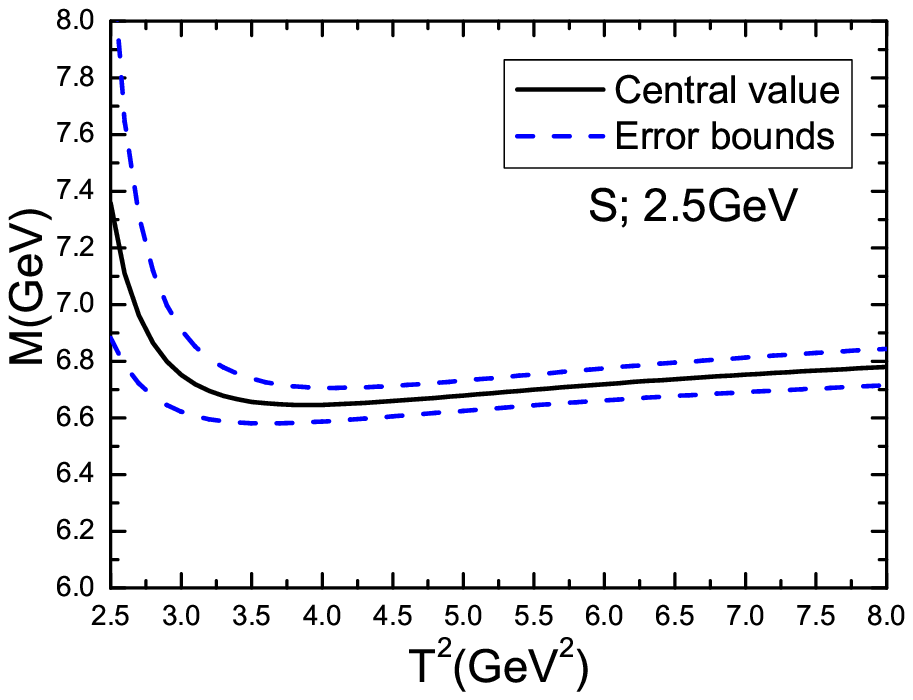}
\includegraphics[totalheight=6cm,width=7cm]{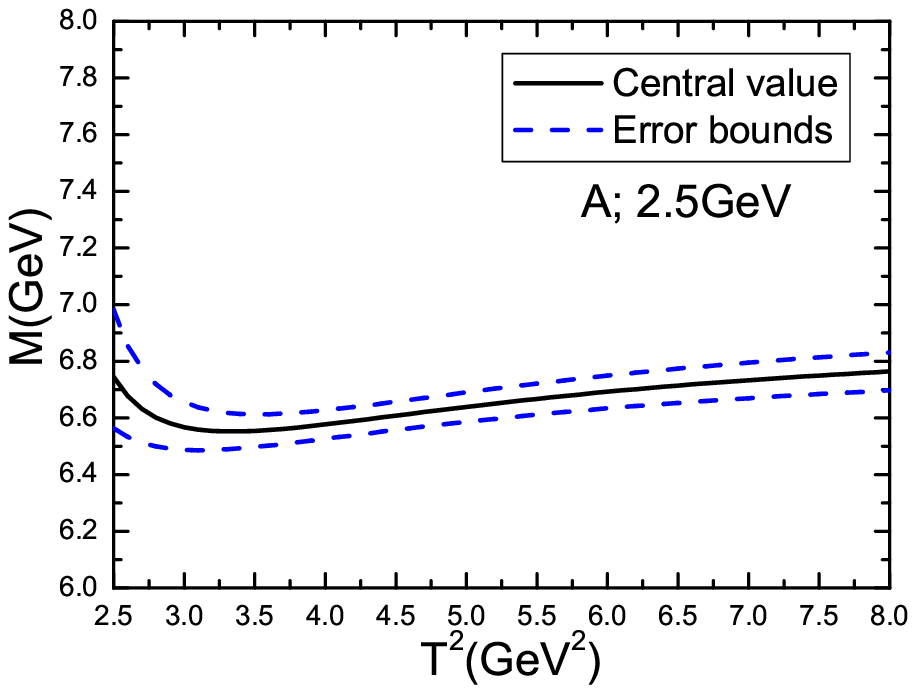}
\includegraphics[totalheight=6cm,width=7cm]{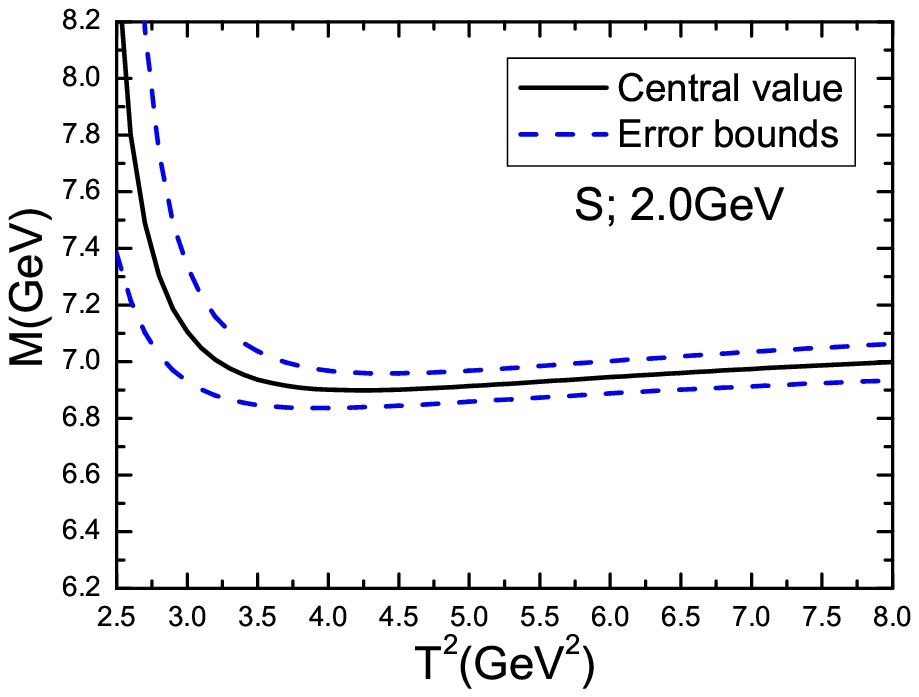}
\includegraphics[totalheight=6cm,width=7cm]{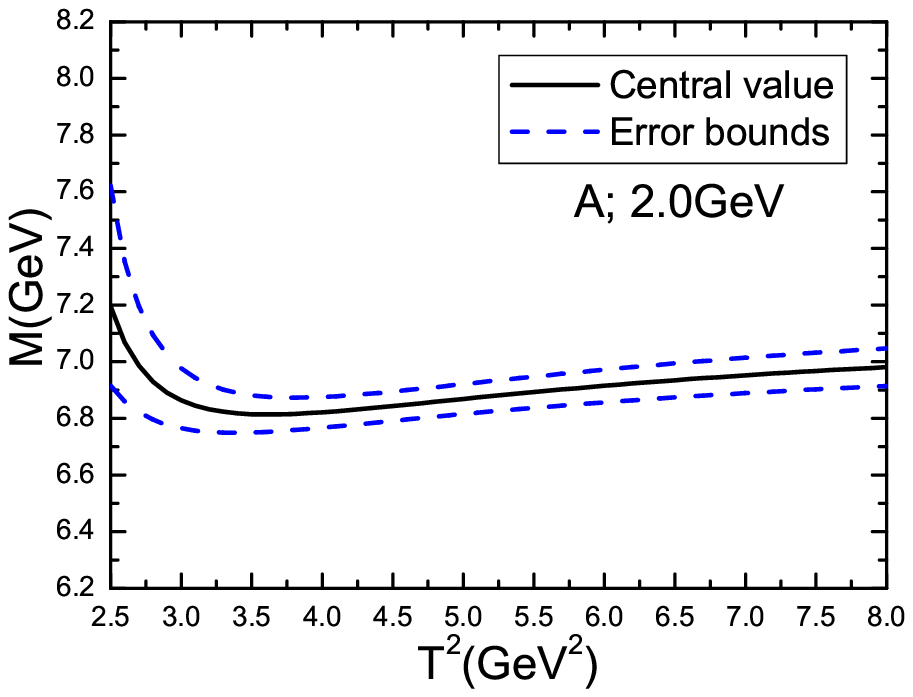}
\includegraphics[totalheight=6cm,width=7cm]{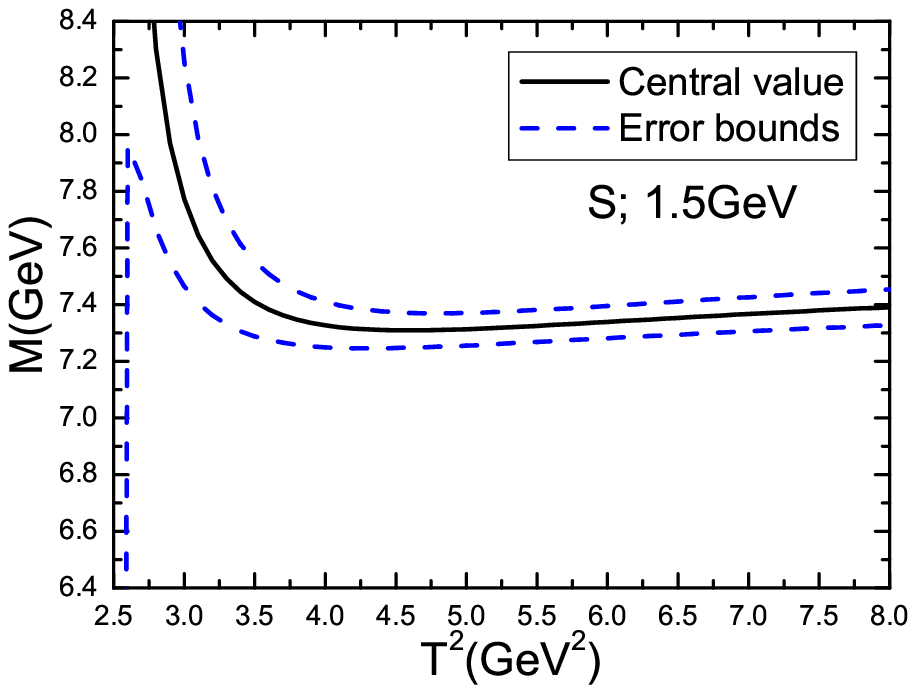}
\includegraphics[totalheight=6cm,width=7cm]{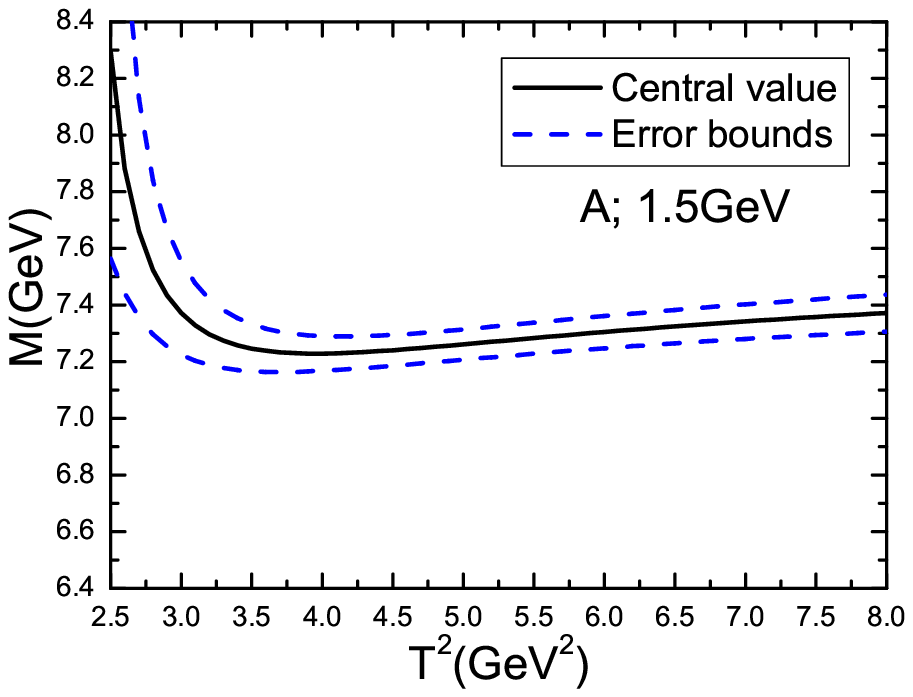}
 \caption{ The masses of the $(bc)_Sq$ and $(bc)_Aq$ baryon states with the $J^P={\frac{1}{2}}^+$  with variations  of the Borel parameters $T^2$, where the $S$ and $A$ denote the
 scalar and axialvector diquarks, respectively, the $2.5\,\rm{GeV}$, $2.0\,\rm{GeV}$ and $1.5\,\rm{GeV}$ denote the energy scales of the QCD spectral densities, $q=u$ and $d$.  }\label{mass-Borel-12}
\end{figure}

In Fig.\ref{mass-Borel-12},  we plot the  masses of the $(bc)_Sq$ and $(bc)_Aq$ baryon states with the spin-parity $J^P={\frac{1}{2}}^+$  with variations  of the Borel parameters $T^2$ at much larger intervals  than the Borel windows shown in Table \ref{Borel-mass-pole} as an example, where $q=u$, $d$. At the Borel windows, there appear very flat platforms, the uncertainties originate from the Borel parameter are very small.

We take  account of  all uncertainties  of the input   parameters,
and obtain  the masses and pole residues of
 the bottom-charm  baryon states, which are shown explicitly in Tables \ref{Borel-mass-pole}-\ref{Borel-mass-pole-2}.

 In calculations, we observe that the masses and pole residues increase monotonously and quickly with the increase of the continuum threshold parameters, the uncertainties originate from the continuum thresholds  $s_0$ are very large and account for  more than half of the total uncertainties, for example, in the case of the $(bc)_Sq$ baryon state with
 the $J^P={\frac{1}{2}}^+$ extracted at the energy scale $\mu=2.0\,\rm{GeV}$, the total uncertainties are $\delta M_{+}=\pm 0.07\,\rm{GeV}$ and $\delta \lambda_{+}=\pm0.09\times 10^{-1}\,\rm{GeV}^{-3}$ respectively, while the uncertainty $\delta \sqrt{s_0}=\pm0.1\,\rm{GeV}$ leads to the uncertainties $\delta M_{+}=\pm 0.04\,\rm{GeV}$ and $\delta \lambda_{+}=\pm0.07\times 10^{-1}\,\rm{GeV}^{-3}$ respectively.
    We determine the continuum threshold parameters $s_0$ by uniform constraints, such as the continuum thresholds  $\sqrt{s_0}\approx M_{+}+0.6 \pm0.1 \,\rm{GeV}$, pole contributions $(50-75)\%$ and  intervals   $T^2_{max}-T^2_{min}=1\,\rm{GeV}^2$ to acquire reliable predictions, where the  $T^2_{max}$ and $T^2_{min}$ stand for the maximum and minimum values of the Borel parameters, respectively.

 From  Tables \ref{Borel-mass-pole}-\ref{Borel-mass-pole-2}, we can see explicitly that the masses $M_{+}\approx 7.3\,\rm{GeV}$ for the $(bc)_Sq$, $(bc)_Aq$, $(qc)_Sb$ and $(qc)_Ab$ baryon states extracted at the energy scale $\mu=1.5\,\rm{GeV}$, which coincide with that of the ground state axialvector bottom-charm tetraquark states,  $M_{Z_{\bar{b}c}(1^{+-})}=7.30\pm0.08\,\rm{GeV}$ and $M_{Z_{\bar{b}c}(1^{++})}=7.31\pm0.08\,\rm{GeV}$ \cite{WZG-EPL-Bc}. It is odd that the three-quark and four-quark bottom-charm hadrons  have the same ground state masses, we discard the values extracted at the energy scale $\mu=1.5\,\rm{GeV}$, just as what we expect.

\begin{table}
\begin{center}
\begin{tabular}{|c|c|c|c|c|c|c|c|}\hline\hline
           &$J^P$              &$\mu(\rm GeV)$   &$T^2(\rm{GeV}^2)$  &$\sqrt{s_0} (\rm{GeV})$  &pole         &$M(\rm{GeV})$  &$\lambda (10^{-1}\rm{GeV}^3)$   \\  \hline
$(bc)_Sq$  &${\frac{1}{2}}^+$  &$2.5$            &$5.0-6.0$          &$7.30\pm0.10$            &$(51-74)\%$  &$6.70\pm0.08$  &$0.75\pm0.08 $    \\ \hline
$(bc)_Aq$  &${\frac{1}{2}}^+$  &$2.5$            &$5.0-6.0$          &$7.30\pm0.10$            &$(51-74)\%$  &$6.67\pm0.08$  &$1.29\pm0.17 $    \\ \hline
$(bc)_Aq$  &${\frac{3}{2}}^+$  &$2.5$            &$5.1-6.1$          &$7.30\pm0.10$            &$(51-73)\%$  &$6.70\pm0.08$  &$0.73\pm0.08 $    \\ \hline

$(bc)_Ss$  &${\frac{1}{2}}^+$  &$2.5$            &$5.4-6.4$          &$7.45\pm0.10$            &$(52-73)\%$  &$6.82\pm0.08$  &$0.90\pm0.11 $    \\ \hline
$(bc)_As$  &${\frac{1}{2}}^+$  &$2.5$            &$5.3-6.3$          &$7.45\pm0.10$            &$(52-74)\%$  &$6.79\pm0.09$  &$1.55\pm0.20 $    \\ \hline
$(bc)_As$  &${\frac{3}{2}}^+$  &$2.5$            &$5.5-6.5$          &$7.45\pm0.10$            &$(51-73)\%$  &$6.82\pm0.08$  &$0.88\pm0.11 $    \\ \hline

$(bc)_Sq$  &${\frac{1}{2}}^+$  &$2.0$            &$5.0-6.0$          &$7.50\pm0.10$            &$(51-74)\%$  &$6.93\pm0.07$  &$0.73\pm0.09 $    \\ \hline
$(bc)_Aq$  &${\frac{1}{2}}^+$  &$2.0$            &$5.0-6.0$          &$7.50\pm0.10$            &$(50-74)\%$  &$6.89\pm0.08$  &$1.25\pm0.16 $    \\ \hline
$(bc)_Aq$  &${\frac{3}{2}}^+$  &$2.0$            &$5.1-6.1$          &$7.50\pm0.10$            &$(50-73)\%$  &$6.93\pm0.07$  &$0.71\pm0.09 $    \\ \hline

$(bc)_Ss$  &${\frac{1}{2}}^+$  &$2.0$            &$5.4-6.4$          &$7.65\pm0.10$            &$(52-73)\%$  &$7.04\pm0.08$  &$0.88\pm0.11 $    \\ \hline
$(bc)_As$  &${\frac{1}{2}}^+$  &$2.0$            &$5.3-6.3$          &$7.65\pm0.10$            &$(52-74)\%$  &$7.01\pm0.08$  &$1.51\pm0.20 $    \\ \hline
$(bc)_As$  &${\frac{3}{2}}^+$  &$2.0$            &$5.5-6.5$          &$7.65\pm0.10$            &$(51-73)\%$  &$7.04\pm0.08$  &$0.86\pm0.11 $    \\ \hline

$(bc)_Sq$  &${\frac{1}{2}}^+$  &$1.5$            &$5.3-6.3$          &$7.90\pm0.10$            &$(51-74)\%$  &$7.33\pm0.07$  &$0.77\pm0.09 $    \\ \hline
$(bc)_Aq$  &${\frac{1}{2}}^+$  &$1.5$            &$5.3-6.3$          &$7.90\pm0.10$            &$(51-73)\%$  &$7.30\pm0.08$  &$1.32\pm0.17 $    \\ \hline
$(bc)_Aq$  &${\frac{3}{2}}^+$  &$1.5$            &$5.4-6.4$          &$7.90\pm0.10$            &$(51-73)\%$  &$7.33\pm0.07$  &$0.75\pm0.09 $    \\ \hline

$(bc)_Ss$  &${\frac{1}{2}}^+$  &$1.5$            &$5.7-6.7$          &$8.05\pm0.10$            &$(52-73)\%$  &$7.44\pm0.08$  &$0.93\pm0.12 $    \\ \hline
$(bc)_As$  &${\frac{1}{2}}^+$  &$1.5$            &$5.7-6.7$          &$8.05\pm0.10$            &$(51-73)\%$  &$7.42\pm0.09$  &$1.60\pm0.21 $    \\ \hline
$(bc)_As$  &${\frac{3}{2}}^+$  &$1.5$            &$5.8-6.8$          &$8.05\pm0.10$            &$(52-73)\%$  &$7.44\pm0.08$  &$0.91\pm0.11 $    \\ \hline

  \hline
\end{tabular}
\end{center}
\caption{ The energy scales $\mu$, Borel parameters $T^2$, continuum threshold parameters $s_0$,
 pole contributions, mass and pole residues for the bottom-charm baryon states, where $q=u$ and $d$.}\label{Borel-mass-pole}
\end{table}

\begin{table}
\begin{center}
\begin{tabular}{|c|c|c|c|c|c|c|c|c|}\hline\hline
           &$J^P$              &$\mu(\rm GeV)$  &$T^2 (\rm{GeV}^2)$  &$\sqrt{s_0}(\rm GeV) $  &pole       &$M (\rm{GeV})$    & $\lambda (10^{-1}\rm{GeV^3}) $  \\ \hline

$(qc)_Sb$  &${\frac{1}{2}}^+$   & 2.5       & $4.8-5.8$          &$7.30\pm0.10$           &$(51-74)\%$    & $6.66\pm0.09$    & $0.75\pm0.07$        \\ \hline
$(qc)_Ab$  &${\frac{1}{2}}^+$   & 2.5       & $4.7-5.7$          &$7.30\pm0.10$           &$(50-74)\%$    & $6.70\pm0.09$    & $1.31\pm0.16$         \\ \hline
$(qc)_Ab$  &${\frac{3}{2}}^+$   & 2.5       & $4.8-5.8$          &$7.30\pm0.10$           &$(50-74)\%$    & $6.71\pm0.08$    & $0.75\pm0.07$    \\ \hline

$(sc)_Sb$  &${\frac{1}{2}}^+$   & 2.5       & $5.2-6.2$          &$7.45\pm0.10$           &$(51-74)\%$    & $6.79\pm0.09$    & $0.91\pm0.09$    \\ \hline
$(sc)_Ab$  &${\frac{1}{2}}^+$   & 2.5       & $5.1-6.1$          &$7.45\pm0.10$           &$(51-74)\%$    & $6.82\pm0.10$    & $1.58\pm0.19$    \\ \hline
$(sc)_Ab$  &${\frac{3}{2}}^+$   & 2.5       & $5.2-6.2$          &$7.45\pm0.10$           &$(52-74)\%$    & $6.83\pm0.08$    & $0.89\pm0.09$    \\ \hline

$(qc)_Sb$  &${\frac{1}{2}}^+$   & 2.0       & $4.8-5.8$          &$7.50\pm0.10$           &$(51-74)\%$    & $6.89\pm0.09$    & $0.73\pm0.07$    \\ \hline
$(qc)_Ab$  &${\frac{1}{2}}^+$   & 2.0       & $4.7-5.7$          &$7.50\pm0.10$           &$(50-74)\%$    & $6.93\pm0.07$    & $1.27\pm0.16$    \\ \hline
$(qc)_Ab$  &${\frac{3}{2}}^+$   & 2.0       & $4.8-5.8$          &$7.50\pm0.10$           &$(50-74)\%$    & $6.94\pm0.09$    & $0.73\pm0.07$    \\ \hline

$(sc)_Sb$  &${\frac{1}{2}}^+$   & 2.0       & $5.2-6.2$          &$7.65\pm0.10$           &$(51-74)\%$    & $7.01\pm0.09$    & $0.88\pm0.09$       \\ \hline
$(sc)_Ab$  &${\frac{1}{2}}^+$   & 2.0       & $5.1-6.1$          &$7.65\pm0.10$           &$(51-74)\%$    & $7.04\pm0.08$    & $1.57\pm0.19$      \\ \hline
$(sc)_Ab$  &${\frac{3}{2}}^+$   & 2.0       & $5.2-6.2$          &$7.65\pm0.10$           &$(51-73)\%$    & $7.05\pm0.09$    & $0.89\pm0.10$    \\ \hline

$(qc)_Sb$  &${\frac{1}{2}}^+$   & 1.5       & $5.1-6.1$          &$7.90\pm0.10$           &$(51-74)\%$    & $7.29\pm0.09$    & $0.75\pm0.09$     \\ \hline
$(qc)_Ab$  &${\frac{1}{2}}^+$   & 1.5       & $5.0-6.0$          &$7.90\pm0.10$           &$(51-74)\%$    & $7.33\pm0.09$    & $1.35\pm0.16$     \\ \hline
$(qc)_Ab$  &${\frac{3}{2}}^+$   & 1.5       & $5.1-6.1$          &$7.90\pm0.10$           &$(50-73)\%$    & $7.34\pm0.09$    & $0.77\pm0.07$     \\ \hline

$(sc)_Sb$  &${\frac{1}{2}}^+$   & 1.5       & $5.5-6.5$          &$8.05\pm0.10$           &$(52-74)\%$    & $7.41\pm0.09$    & $0.92\pm0.10$     \\ \hline
$(sc)_Ab$  &${\frac{1}{2}}^+$   & 1.5       & $5.4-6.4$          &$8.05\pm0.10$           &$(51-74)\%$    & $7.44\pm0.08$    & $1.63\pm0.19$     \\ \hline
$(sc)_Ab$  &${\frac{3}{2}}^+$   & 1.5       & $5.5-6.5$          &$8.05\pm0.10$           &$(51-73)\%$    & $7.45\pm0.09$    & $0.93\pm0.08$     \\ \hline\hline
\end{tabular}
\end{center}
\caption{ The energy scales $\mu$, Borel parameters $T^2$, continuum threshold parameters $s_0$,
 pole contributions, mass and pole residues for the bottom-charm baryon states, where $q=u$ and $d$.}\label{Borel-mass-pole-2}
\end{table}

In the non-relativistic QCD sum rules, Kiselev and  Likhoded obtain the values $M_{(cc)_Aq} = 3.47 \pm 0.05\,\rm{ GeV}$, $M_{(bc)_Aq} = 6.80 \pm 0.05\,\rm{ GeV}$, $M_{(bb)_Aq} = 10.07 \pm 0.09\,\rm{ GeV}$
for the doubly-heavy baryon states with the spin-parity  $J^P={\frac{1}{2}}^+$ \cite{NRQCDSR-Kiselev}. It is obvious that the prediction $M_{(cc)_Aq} = 3.47 \pm 0.05\,\rm{ GeV}$ in Ref.\cite{NRQCDSR-Kiselev} lies much below the LHCb experimental data.

In the full QCD sum rules, Zhang and  Huang obtain the values $M_{(bc)_Sq} = 6.95 \pm 0.08 \,\rm{ GeV}$,  $M_{(bc)_Aq} = 6.75 \pm 0.05\,\rm{ GeV}$ and $M_{(bc)_Aq} = 8.00 \pm 0.26 \,\rm{ GeV}$ for the bottom-charm baryon states with the spin-parity $J^P={\frac{1}{2}}^+$, ${\frac{1}{2}}^+$ and ${\frac{3}{2}}^+$, respectively \cite{ZhangJR-Doubly}.
It is difficult to compare the present predictions to that in Ref.\cite{ZhangJR-Doubly}, as we choose different scheme in dealing with the QCD sum rules. In Ref.\cite{ZhangJR-Doubly}, Zhang and Huang do not take account of the energy scale dependence of the QCD sum rules and do not separate the contributions of the positive-parity and negative-parity baryon states, furthermore, their Borel  platforms are not flat enough.  In Ref.\cite{WZG-EPJC-cc-baryon}, we investigate the $QQq$-type doubly-heavy baryon states with $q=u$, $d$, $s$, and choose the energy scales $\mu=1\,\rm{GeV}$ and $2.2\,\rm{GeV}$ in the charm and bottom sections respectively, and reach the pole contributions $(65-85)\%$ and $(55-75)\%$ for the $ccq$ and $bbq$ baryon states, respectively, which are compatible with the present pole contributions $(50-75)\%$. More importantly,  we can reproduce the experimental value of the $\Xi_{cc}^{++}$ from the LHCb collaboration \cite{LHCb-Xicc}.

\begin{figure}
 \centering
 \includegraphics[totalheight=6cm,width=7cm]{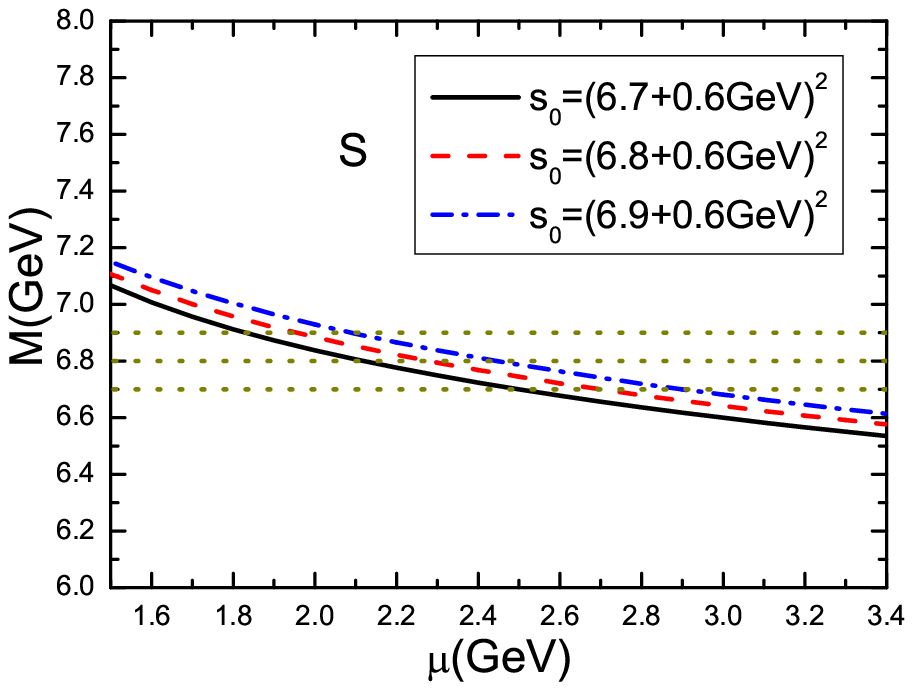}
\includegraphics[totalheight=6cm,width=7cm]{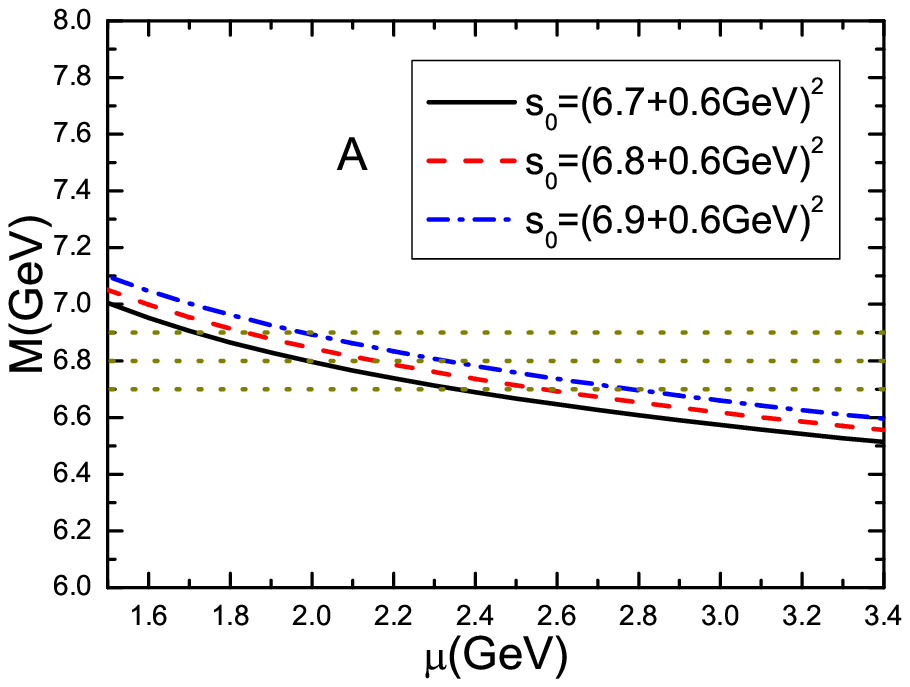}
 \caption{ The masses of the $(bc)_Sq$ and $(bc)_Aq$ baryon states with the $J^P={\frac{1}{2}}^+$  with variations  of the energy scales $\mu$ for the Borel parameters $T^2=5.5\,\rm{GeV}^2$, where the $S$ and $A$ denote the
 scalar and axialvector diquarks, respectively, $q=u$ and $d$.  }\label{mass-s0-mu}
\end{figure}

For the bottom-charm mesons, the mass gaps $M_{B_c(\rm 2S)}-M_{B_c}=596 \,\rm{MeV}$ and
$M_{B_c^*(\rm 2S)}-M_{B^*_c}=567\,\rm{MeV}$ from the CMS collaboration \cite{CMS-BcV}, and $M_{B_c^*(\rm 2S)}-M_{B^*_c}=566 \,\rm{MeV}$ from the LHCb collaboration   \cite{LHCb-Bc-1904}. Analogously, we take the constraint $\sqrt{s_0}=M_{gr}+0.6\,\rm{GeV}$ for the ground state (gr) bottom-charm baryon states. In Fig.\ref{mass-s0-mu}, we plot the predicted masses of the $(bc)_Sq$ and $(bc)_Aq$ baryon states with the spin-parity $J^P={\frac{1}{2}}^+$ with variations  of the energy scales $\mu$ for the Borel parameters $T^2=5.5\,\rm{GeV}^2$ as an example. From the figure, we can see that the constraint $\sqrt{s_0}=M_{gr}+0.6\,\rm{GeV}$ plays an important role in selecting the pertinent energy scales $\mu$, the largest  energy scales $\mu$ should not exceed $2.8\,\rm{GeV}$, even the energy scale $\mu=2.6\,\rm{GeV}$, to avoid contaminations of the continuum states due to the large energy gaps $\sqrt{s_0}-M_{gr}$; on the other hand, the lower bound of the energy scales is about $2.0\,\rm{GeV}$, an even lower energy scale would fail to take account of the ground state contributions fully  due to the small energy gaps $\sqrt{s_0}-M_{gr}$. From the Particle Data Group, we take the pole mass  of the $b$ quark, $\hat{m}_b=4.78\pm0.06\,\rm{ GeV}$ \cite{PDG}. The central value of the $\overline{MS}$ mass  $m_b(\mu=2\,\rm{GeV})=4.77\,\rm{GeV}$ happens to  coincide with the pole mass  $4.78\,\rm{GeV}$, so the energy scale $\mu=2.0\,\rm{GeV}$ is a typical energy scale (or the lowest energy scale) in the QCD sum rules for the hadron masses  involving the $b$ quark. Naively, we expect to choose the lowest energy scales of the QCD spectral densities so as to manifest the contributions of the vacuum condensates therefore manifest the ground state contributions  to acquire stable QCD sum rules.  The predicted masses vary with the energy scales of the QCD spectral densities significantly, we prefer the predictions in Tables \ref{Borel-mass-pole}-\ref{Borel-mass-pole-2} with the energy scale $\mu=2.0\,\rm{GeV}$, while the values extracted at the energy scale $\mu=2.5\,\rm{GeV}$ can not be  excluded (or are  also acceptable) , whether or not such a preference works, we can confront the predictions with  the experimental data in the future (to select the best energy scales of the QCD spectral densities).

In the QCD sum rules, we usually choose the local currents, in fact, the physical baryon states have finite sizes, therefore the bottom-charm baryons  can be symbolically  written as  $\varepsilon^{ijk}b_i(x)c_j(x+\epsilon)q_k(x+\theta)$, where the $\epsilon^\mu=(0,\vec{\epsilon})$ and $\theta^\mu=(0,\vec{\theta})$ are finite four-vectors in the coordinate space. In the heavy quark limit,
\begin{eqnarray}
\langle \vec{\epsilon}\rangle &\sim & \frac{1}{\widetilde{m}_Q v}\, , \nonumber\\
\langle \vec{\theta}\rangle &\sim & \frac{1}{\Lambda_{QCD}}\, , \nonumber\\
\langle \vec{\theta}-\vec{\epsilon}\rangle &\sim & \frac{1}{\Lambda_{QCD}}\, ,
\end{eqnarray}
where $\widetilde{m}_Q=\frac{m_bm_c}{m_b+m_c}$, the $v$ is the heavy quark velocity,
\begin{eqnarray}
\frac{\langle \vec{\epsilon}\rangle}{\langle \vec{\theta}\rangle} &\sim&\frac{\Lambda_{QCD}}{\widetilde{m}_Q v}\, ,\nonumber\\
\frac{\langle \vec{\epsilon}\rangle}{\langle \vec{\theta}-\vec{\epsilon}\rangle} &\sim&\frac{\Lambda_{QCD}}{\widetilde{m}_Q v}\, ,
\end{eqnarray}
the distance between the $b$ and $c$ quarks is much smaller than that between the $b/c$ and $q$ quarks \cite{FKGuo-bc}. In the infinite heavy quark limit,
the $b$ and $c$ quarks are static and serve as a static well potential in the color anti-triplet, and attract the light quark $q$ to form a baryon, just like the  hydrogen atom. It is better to choose the currents $J^S(x)$, $J^A(x)$ and $J^A_\mu(x)$, where the $b$ and $c$ quarks formulate diquark correlations in the color anti-triplet, and we prefer the configurations $(bc)_{S/A}q$.

We should bear in mind that the $c$ quark is not heavy enough and we take the local limit $\vec{\epsilon}\to 0$ and $\vec{\theta}\to 0$ in the QCD sum rules. If we perform Fierz transformation for the currents $J^S(x)$, $J^A(x)$ and $J^A_\mu(x)$, those currents can be changed into superpositions of serval currents,  where the $c$ and $q$ quarks form diquark correlations in the color anti-triplet,
\begin{eqnarray}\label{Fierz-JS}
 J^{S}&=&\frac{1}{4}\varepsilon^{ijk}q^T_i C\gamma_5 c_{j}\,b_{k}-\frac{1}{4}\varepsilon^{ijk}q^T_i C\gamma^\alpha c_{j} \,\gamma^\alpha \gamma_5 b_{k}+\frac{1}{4}\varepsilon^{ijk}q^T_i C c_{j}\,\gamma_5b_{k}-\frac{1}{4}\varepsilon^{ijk}q^T_i C\gamma^\alpha \gamma_5c_{j}\,\gamma^\alpha b_{k} \nonumber \\
&&-\frac{1}{8}\varepsilon^{ijk}q^T_i C\sigma^{\alpha\beta} c_{j}\, \sigma_{\alpha\beta} \gamma_5 b_{k} \, ,\nonumber\\
&=&\frac{1}{4}\eta^S+\frac{1}{4}\eta^A+\frac{1}{4}\varepsilon^{ijk}q^T_i C c_{j}\,\gamma_5b_{k}-\frac{1}{4}\varepsilon^{ijk}q^T_i C\gamma^\alpha \gamma_5c_{j}\,\gamma^\alpha b_{k} -\frac{1}{8}\varepsilon^{ijk}q^T_i C\sigma^{\alpha\beta} c_{j}\, \sigma_{\alpha\beta} \gamma_5 b_{k} \, ,
\end{eqnarray}
\begin{eqnarray}\label{Fierz-JA}
J^{A}&=&\varepsilon^{ijk}q^T_i C\gamma_5 c_{j}\, b_{k}+\frac{1}{2}\varepsilon^{ijk}q^T_i C\gamma_\mu c_{j}\, \gamma^\mu \gamma_5 b_{k}-\varepsilon^{ijk}q^T_i C c_{j}\, \gamma_5 b_{k}-\frac{1}{2}\varepsilon^{ijk}q^T_i C\gamma_\mu \gamma_5 c_{j}\,  \gamma^\mu b_{k} \, ,\nonumber \\
&=&\eta^S-\frac{1}{2}\eta^A-\varepsilon^{ijk}q^T_i C c_{j}\, \gamma_5 b_{k}-\frac{1}{2}\varepsilon^{ijk}q^T_i C\gamma_\mu \gamma_5 c_{j}\,  \gamma^\mu b_{k} \, ,
\end{eqnarray}
\begin{eqnarray}\label{Fierz-JA-mu}
J_\mu^{A}&=&\frac{1}{4}\varepsilon^{ijk}q^T_i C\gamma_\mu c_{j}\, b_{k}-\frac{1}{4}\varepsilon^{ijk}q^T_i C\gamma_5 c_{j}\,\gamma_5 \gamma_\mu b_{k}-\frac{1}{4}\varepsilon^{ijk}q^T_i C c_{j}\,\gamma_\mu b_{k}-\frac{1}{4}\varepsilon^{ijk}q^T_i C\gamma_\mu \gamma_5 c_{j}\,\gamma_5 b_{k}\nonumber \\
&&+\frac{i}{4}\varepsilon^{ijk}q^T_i C\gamma^\alpha \gamma_5 c_{j}\,\sigma_{\mu\alpha}\gamma_5 b_{k}
-\frac{i}{4}\varepsilon^{ijk}q^T_i C\gamma^\alpha c_{j}\,\sigma_{\mu\alpha} b_{k}+\frac{1}{8}\varepsilon^{ijk}q^T_i C\sigma^{\alpha\beta} c_{j}\, \sigma_{\alpha\beta} \gamma_\mu b_{k} \, , \nonumber \\
&=&\frac{1}{4}\eta_\mu^A-\frac{1}{4}\varepsilon^{ijk}q^T_i C\gamma_5 c_{j}\,\gamma_5 \gamma_\mu b_{k}-\frac{1}{4}\varepsilon^{ijk}q^T_i C c_{j}\,\gamma_\mu b_{k}-\frac{1}{4}\varepsilon^{ijk}q^T_i C\gamma_\mu \gamma_5 c_{j}\,\gamma_5 b_{k}\nonumber \\
&&+\frac{i}{4}\varepsilon^{ijk}q^T_i C\gamma^\alpha \gamma_5 c_{j}\,\sigma_{\mu\alpha}\gamma_5 b_{k}
-\frac{i}{4}\varepsilon^{ijk}q^T_i C\gamma^\alpha c_{j}\,\sigma_{\mu\alpha} b_{k}+\frac{1}{8}\varepsilon^{ijk}q^T_i C\sigma^{\alpha\beta} c_{j}\, \sigma_{\alpha\beta} \gamma_\mu b_{k} \, .
\end{eqnarray}
From Eqs.\eqref{Fierz-JS}-\eqref{Fierz-JA}, we can see that $J^{S}=\frac{1}{4}\eta^S+\frac{1}{4}\eta^A+\cdots$ and $J^{S}=\eta^S-\frac{1}{2}\eta^A+\cdots$, on the other hand, from Table \ref{Borel-mass-pole-2}, we can see that the central values of the $(qc)_{S/A}b$ baryon states with the $J^P={\frac{1}{2}}^+$ have the relation $M_A> M_S$, it is consistent with the predictions in Table \ref{Borel-mass-pole}, where the  central values of the $(bc)_{S/A}q$ baryon states with the $J^P={\frac{1}{2}}^+$ have the relation $M_A< M_S$ due to the mixing coefficients $\frac{1}{4}$, $\frac{1}{4}$, $1$, $\frac{1}{2}$, which serves as a crosscheck of the reliability of the QCD sum rules.

\section{Conclusion}
In this article, we  construct the  $(bc)_Sq$,  $(bc)_Aq$, $(qc)_Sb$ and $(qc)_Ab$ type diquark-quark currents to  explore  the ground state mass spectrum of the bottom-charm baryon states with the positive-parity in the framework of
 the QCD sum rules in a comprehensive way by carrying out the operator product expansion   up to   the vacuum condensates of dimension $7$.
In calculations,   we distinguish  the contributions of the positive-parity and negative-parity baryon states  unambiguously to avoid contaminations,
and  investigate the energy scale dependence of the QCD sum rules in details and observe that the acceptable energy scales are $\mu=2.0\sim 2.8\,\rm{GeV}$, and extract the  masses and pole residues of the bottom-charm baryon states at the typical  energy scales $\mu=2.5\,\rm{GeV}$, $2.0\,\rm{GeV}$ and $1.5\,\rm{GeV}$, respectively.
In the heavy quark limit, it is better to choose the configurations $(bc)_Sq$ and  $(bc)_Aq$ due to the small distance between the $b$ and $c$ quarks, the configurations $(bc)_Sq$ and  $(bc)_Aq$ relate to the configurations $(qc)_Sb$ and $(qc)_Ab$ through Fierz re-arrangements, the predicted masses for the two different  configurations serve as a crosscheck of the reliability of the QCD sum rules.
 The preferred values are extracted at the energy scale  $\mu=2.0\,\rm{GeV}$, while the values extracted at the energy scale $\mu=2.5\,\rm{GeV}$ are also acceptable (or cannot be excluded),  the present predictions can be confronted to the experimental data in the future (to select the best energy scales of the QCD spectral densities) and shed light on the QCD low energy dynamics.

\section*{Acknowledgements}
This  work is supported by National Natural Science Foundation, Grant Number 12175068.

\end{document}